\documentclass[manuscript,screen]{acmart}

\usepackage{booktabs} 
\usepackage{amsmath}

\usepackage{algorithmic}
\usepackage{graphicx}
\usepackage{algorithm,algorithmic}

\usepackage{longtable}

\def\BibTeX{{\rm B\kern-.05em{\sc i\kern-.025em b}\kern-.08em
    T\kern-.1667em\lower.7ex\hbox{E}\kern-.125emX}}

\title{Enhancing Online Support Group Formation Using Topic Modeling Techniques}

\author{Pronob Kumar Barman}
\affiliation{%
  \department{Department of Information Systems}
  \institution{University of Maryland, Baltimore County}
  \city{Baltimore}
  \state{Maryland}
  \country{USA}
}
\email{pbarman1@umbc.edu}

\author{Tera L. Reynolds}
\affiliation{%
  \department{Department of Information Systems}
  \institution{University of Maryland, Baltimore County}
  \city{Baltimore}
  \state{Maryland}
  \country{USA}
}
\email{reynoter@umbc.edu}

\author{James Foulds}
\affiliation{%
  \department{Department of Information Systems}
  \institution{University of Maryland, Baltimore County}
  \city{Baltimore}
  \state{Maryland}
  \country{USA}
}
\email{jfoulds@umbc.edu}

\begin{document}
\begin{abstract}
Online health communities (OHCs) are vital for fostering peer support and improving health outcomes. Support groups within these platforms can provide more personalized and cohesive peer support, yet traditional support group formation methods face challenges related to scalability, static categorization, and insufficient personalization. To overcome these limitations, we propose two novel 
machine learning models for automated support group formation: the Group-specific Dirichlet Multinomial Regression (gDMR) and the Group-specific Structured Topic Model (gSTM). These models integrate user-generated textual content, demographic profiles, and interaction data represented through node embeddings derived from user networks to systematically automate personalized, semantically coherent support group formation.

We evaluate the models on a large-scale dataset from MedHelp.org, comprising over 2 million user posts. Both 
models substantially outperform baseline methods—including LDA, DMR, and STM—in predictive accuracy (held-out log-likelihood), semantic coherence (UMass metric), and internal group consistency. The gDMR model yields group covariates that facilitate practical implementation by leveraging relational patterns from network structures and demographic data. In contrast, gSTM emphasizes sparsity constraints to generate more distinct and thematically specific groups. Qualitative analysis further validates the alignment between model-generated groups and manually coded themes, showing the practical relevance of the models in informing groups that address diverse health concerns such as chronic illness management, diagnostic uncertainty, and mental health. By reducing reliance on manual curation, these frameworks provide scalable solutions that enhance peer interactions within OHCs, with implications for patient engagement, community resilience, and health outcomes. 
\end{abstract}

\keywords{automated support groups, health informatics, node embeddings, online health forums, personalized support systems, semantic coherence, topic modeling}
\maketitle

\section{Introduction}
\label{sec:introduction}

Online health communities (OHCs), such as \textit{PatientsLikeMe} and \textit{MedHelp}, have emerged as vital platforms enabling patients to access emotional, informational, and social support, particularly in managing chronic illnesses and mental health conditions \cite{wang2018, necaise2024peer, zhao2022}. The proliferation of digital health forums, ranging from disease-specific discussion boards to expansive social media groups, underscores their significance as accessible resources for patients seeking peer interaction and shared understanding \cite{effectivenessSupportGroups, MayoClinicSupportGroups, greene2011}. Empirical evidence consistently demonstrates that active engagement within these communities correlates positively with improved emotional well-being, better self-management practices, and enhanced health outcomes \cite{liu2020, kiser2001social, barak2008}.

Despite these benefits, significant challenges persist in harnessing the full therapeutic potential of OHCs. Organically formed communities often experience rapid growth, resulting in large, heterogeneous groups where active engagement remains limited and lurking behavior predominates \cite{kauf2020}. The sheer volume of participants complicates users’ ability to identify groups aligning precisely with their emotional and informational needs, thereby limiting the efficacy of peer support \cite{hwang2021}. Traditional manual approaches to forming smaller, structured support groups often rely on simple heuristics such as self-selection, symptom-based categorization, or clinician-led grouping. These approaches typically overlook existing user connections, interaction patterns, and complex social relationships within broader communities, leading to suboptimal group cohesion and limited personalization \cite{eysenbach2004, breuer2015}. Consequently, patients may not fully reap the benefits intended by these support environments, such as increased treatment adherence and improved psychological resilience \cite{barak2008, kiser2001social}.

To address these limitations, computational methodologies for automating support group formation have attracted considerable interest. Graph-based methods utilizing person-generated health data (PGHD) have shown promise in constructing semantically coherent and personalized groups by leveraging latent user interaction patterns \cite{hartzler2016leveraging, fang2022matching, fang2022graph}. Furthermore, hybrid approaches integrating textual content, demographic data, and relational patterns have significantly advanced the effectiveness and scalability of group formation techniques \cite{liang2021learning, sullivan2019exploring, valles2023minimum}. However, critical challenges such as ensuring fairness in group assignments, safeguarding user privacy, and maintaining interpretability in large-scale environments remain largely unresolved \cite{hashmi2022, leung2021, zhao2022privacy}.

Motivated by these persistent gaps, this paper introduces two innovative and complementary models explicitly designed to automate the formation of personalized support groups within OHCs: the \emph{Group-specific Dirichlet Multinomial Regression} (gDMR) model and the \emph{Group-specific Structured Topic Model} (gSTM). The gDMR model extends the foundational Dirichlet Multinomial Regression (DMR) framework \cite{mimno2012topic} by incorporating node embeddings derived from user interaction networks and group-specific parameters. This enables nuanced capture of demographic characteristics and relational context, significantly enhancing the personalization and relevance of group assignments. Complementarily, the gSTM builds upon the Structured Topic Model (STM) \cite{roberts2013structural}, introducing sparsity-inducing priors and structured covariates alongside group-specific deviations to enhance topic coherence and interpretability within groups.

Our empirical evaluations demonstrate that these models substantially outperform baseline approaches, such as standard Latent Dirichlet Allocation (LDA) \cite{blei2003latent}, traditional DMR, and conventional STM, in critical metrics including held-out log-likelihood and topic coherence \cite{mimno2012topic, roberts2013structural}. The integration of relational data in gDMR notably improves the accuracy and personalization of group formation, while gSTM excels in generating semantically rich and interpretable thematic structures, thus supporting meaningful peer interactions. We acknowledge that challenges regarding fairness in group allocation and robust privacy preservation remain partially addressed, highlighting avenues for future research.

This research provides a scalable, data-driven paradigm for automating personalized support group formation, significantly advancing the state-of-the-art in OHC management. By mitigating the limitations inherent to manual group formation methods, the gDMR and gSTM models have the potential to enhance patient engagement, facilitate more cohesive peer support networks, and ultimately contribute to improved health outcomes and more resilient online communities.

\section{Background and Related Work}

\subsection{Role and Challenges of Online Health Communities and Support Groups}

OHCs are broad digital platforms where patients and caregivers engage in discussions, share experiences, and seek emotional and informational support related to a wide range of health concerns \cite{newman2011s, ressler_2012, ziebland2012health}. Examples include disease-specific forums like \textit{Diabetes Daily} or expansive social media groups centered on mental health. These communities often comprise thousands to millions of users and provide open spaces for interaction.

Within these expansive OHCs, \emph{support groups} refer to smaller, more focused sub-communities formed around specific conditions, experiences, or needs. Support groups typically offer a more intimate setting that facilitates deeper connections and tailored peer support, often resulting in improved participant engagement and better health outcomes \cite{eysenbach2004, breuer2015, barman2026understanding}. 

Support groups can form in various ways:  
- \textbf{Manual formation} through user self-selection or facilitation by moderators or healthcare professionals, often based on criteria such as diagnosis or treatment phase;  
- \textbf{Algorithmic or automated formation} leveraging computational techniques that consider user-generated data and interaction patterns.

Manual methods may rely primarily on demographic or diagnostic information, sometimes overlooking complex social interactions and evolving user needs. While straightforward, such methods may produce groups with limited cohesion and inadequate personalization, failing to capture the nuanced social relationships that influence peer support effectiveness.

\subsection{Computational Approaches to Support Group Formation}

Recent computational approaches aim to automate support group formation by exploiting rich data sources and advanced modeling techniques. Approaches such as those utilizing PGHD—including user posts, activity logs, and symptom reports—have shown promise in aligning peer matches based on shared health experiences, thereby promoting meaningful connections \cite{hartzler2016leveraging, yeo2023enhancing}. 

Graph-based algorithms, particularly those employing embedding techniques like Node2Vec \cite{grover2016node2vec}, capture latent interaction structures between users, enabling clustering methods that consider both social ties and content similarities \cite{fang2022matching, fang2022graph}. Hybrid frameworks that combine textual content analysis, demographic attributes, and interaction data have emerged as comprehensive solutions addressing the dual challenges of scalability and personalization in group formation \cite{liang2021learning, sullivan2019exploring, zhao2022}.

\subsection{Current Limitations and Gaps}

Despite promising progress, several critical limitations remain. Many existing methodologies rely on \emph{static categorizations}, where users are assigned to groups based on fixed attributes or pre-defined labels that do not reflect evolving behaviors, changing health statuses, or shifting social connections \cite{bergtold2024assessment}. This rigidity limits the adaptability of support groups to meet users’ interaction and complex needs.

Moreover, ethical challenges such as bias and fairness in group assignment frequently remain under-addressed, risking the marginalization of vulnerable populations \cite{hashmi2022}. Scalability and interpretability also present ongoing obstacles, especially when applying models to large-scale, continuously growing online communities. Robust privacy protections are essential but often difficult to integrate effectively \cite{leung2021, zhao2022privacy}.

Our work addresses these limitations by developing novel extensions of probabilistic topic modeling frameworks. Topic modeling enables simultaneous extraction of latent thematic structures from textual data while incorporating user features and interaction patterns. This integrated approach facilitates flexible, data-driven group formation that adapts to users’ evolving communication and engagement patterns. Furthermore, topic modeling inherently supports interpretability through semantically coherent topics, making group assignments more transparent and actionable.


\section{Methods}

\subsection{Dataset Overview}

The dataset utilized in this study was sourced from \textbf{MedHelp.org}, a widely known OHC that operated for nearly three decades before ceasing on May 31, 2024 \cite{medhelp_reference}. MedHelp served as a platform where users engaged in discussions on a wide range of health-related topics. The dataset comprises over 2 million user-generated questions and 8 million corresponding answers, contributed by more than 2 million users. These discussions cover an extensive variety of health conditions and concerns, ranging from chronic diseases such as diabetes and asthma to general health advice, including fitness routines and dietary suggestions.

Each post is enriched with user metadata, which includes:

\begin{itemize}
    \item \textbf{Gender and Age}: Facilitates demographic analyses to discern trends in health-related discussions across different user groups.
    \item \textbf{Membership Year}: Enables examination of user engagement patterns over time, offering insights into the evolution of community participation.
    \item \textbf{Location}: Allows for regional analysis of health discussions, aiding in the identification of geographically specific health concerns and trends.
\end{itemize}

The richness of this dataset provides a valuable resource for analyzing user interactions and information dissemination within online health communities. The inclusion of demographic and temporal metadata enhances the potential for conducting comprehensive studies on health communication patterns, user engagement relationships, and the impact of regional factors on health-related discussions.

\subsubsection*{Data Preprocessing} 
We begin by cleaning, standardizing, and tokenizing the textual data to prepare it for analysis. Demographic features are normalized for consistency, and interaction data is used to generate node embeddings, which capture latent relationships between users.

\subsection{Model Overview}

We propose two complementary models, the gDMR and the gSTM, to automate the formation of personalized support groups within online healthcare forums. Both models integrate user-generated content, demographic information, and interaction data to identify nuanced user relationships and generate contextually relevant support groups. While gDMR extends the DMR framework, gSTM builds upon the STM to further enhance flexibility and interpretability in group-level topic modeling.

\subsubsection{Group-specific Dirichlet Multinomial Regression (gDMR)}

The gDMR model builds upon the DMR framework introduced by Mimno et al. \cite{mimno2012topic}. It incorporates group-specific parameters and node embeddings to capture intricate patterns in user behavior, demographics, and textual content, enabling the formation of highly personalized support groups~\cite{Barman2025MedInfo}.

\begin{figure}[t]
        \centering
        \includegraphics[width=\linewidth]{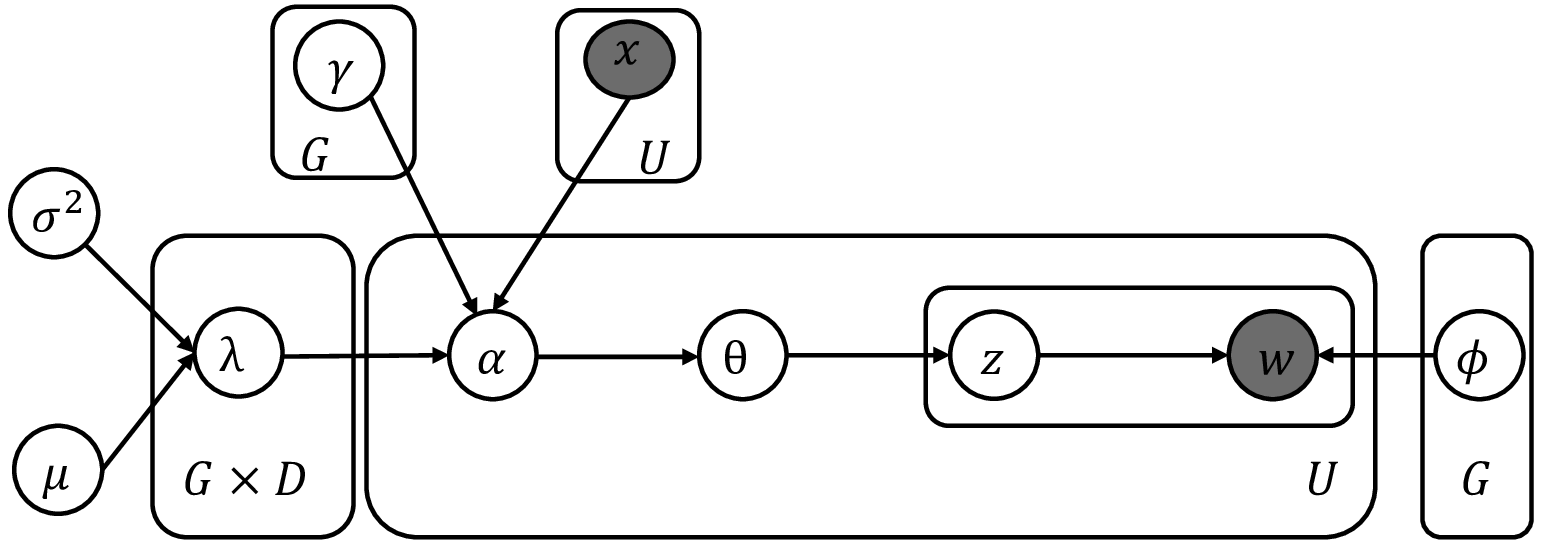}
        \caption{The gDMR Model Plate Diagram.}
        \label{fig:model_gdmr}
\end{figure}

\textbf{Generative Process:}

\begin{enumerate}
    \item For each group \( g \):
    \begin{enumerate}
        \item Draw word distribution \( \phi_g \sim \text{Dirichlet}(\beta) \).
        \item Draw group-specific pseudocounts \( \gamma_g \sim \text{Gamma}(1, 1) \).
        \item Draw regression coefficients \( \lambda_g \sim \mathcal{N}(\mu, \sigma^2I) \).
    \end{enumerate}

    \item For each user \( u \):
    \begin{enumerate}
        \item Compute group membership weight:
        \[
        \alpha_{ug} = \exp({x}_u^T \lambda_g) + \gamma_g.
        \]
        \item Draw group membership proportions \( \theta_u \sim \text{Dirichlet}(\alpha_u) \).
    \end{enumerate}

    \item For each word \( w_{u,n} \) of user \( u \):
    \begin{enumerate}
        \item Draw group assignment \( z_{u,n} \sim \text{Categorical}(\theta_u) \).
        \item Draw word \( w_{u,n} \sim \text{Categorical}(\phi_{z_{u,n}}) \).
    \end{enumerate}
\end{enumerate}

In this model, the regression coefficients \( \lambda_g \) control the influence of user demographic covariates \( x_u \) on group memberships, while the node embeddings capture latent interaction structures. This combination allows gDMR to identify contextually relevant and demographically aligned support groups.

\subsubsection{Group-specific Structured Topic Model (gSTM)}

The gSTM model extends the STM framework~\cite{roberts2013structural} by introducing group-specific deviations, sparsity-inducing priors, and structured covariates to model fine-grained variations in group-level topics. Unlike gDMR, which employs a Dirichlet distribution for group memberships, gSTM leverages a \textbf{Logistic-Normal distribution} to provide greater flexibility in capturing correlations between group memberships.

\begin{figure}[t]
        \centering
        \includegraphics[width=\linewidth]{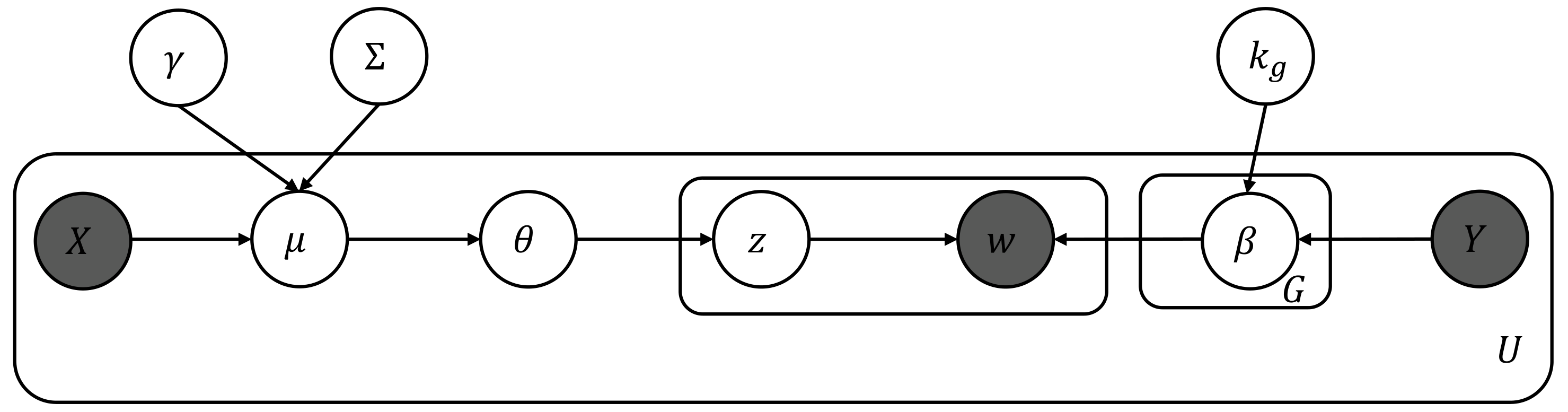}
        \caption{The gSTM Model Plate Diagram.}
        \label{fig:model_gstm}
\end{figure}

\textbf{Generative Process:}

\begin{enumerate}
    \item For each group \( g \):
    \begin{enumerate}
        \item Draw group-specific deviations \( \kappa_{x,w}^g \sim \text{Laplace}(0, \tau_{x,w}^g) \).
        \item Draw sparsity parameter \( \tau_{x,w}^g \sim \text{Gamma}(1, 1) \).
        \item Draw group-specific prior \( \gamma_g \sim \text{Gamma}(1, 1) \).
        \item Draw regression coefficients \( \lambda_g \sim \mathcal{N}(0, \sigma^2I) \).
        \item Compute the group-specific word distribution:
        \[
        \beta_g^w \propto \exp(m_w + \kappa_{x,w}^g).
        \]
    \end{enumerate}
    
    \item For each user \( u \):
    \begin{enumerate}
        \item Compute the group membership mean:
        \[
        \mu_{u,g} = X_u^T \lambda_g + \gamma_g.
        \]
        \item Draw group membership proportions \( \theta_u \sim \text{Logistic Normal}(\mu_u, \Sigma) \).
    \end{enumerate}
    
    \item For each word \( w_{u,n} \) of user \( u \):
    \begin{enumerate}
        \item Draw group assignment \( z_{u,n} \sim \text{Categorical}(\theta_u) \).
        \item Draw word \( w_{u,n} \sim \text{Categorical}(\beta_{z_{u,n}}) \).
    \end{enumerate}
\end{enumerate}

In gSTM, group-specific word distributions \( \beta_g^w \) are modeled using group-level deviations \( \kappa_{x,w}^g \), which are drawn from a sparsity-inducing Laplace prior. This enables gSTM to capture fine-grained group variations while maintaining semantic coherence. The use of a Logistic-Normal distribution for group memberships \( \theta_u \) allows gSTM to account for correlations between group assignments, offering greater flexibility compared to the Dirichlet-based gDMR model.

Both gDMR and gSTM are designed to automate support group formation by leveraging textual, demographic, and interaction data. While gDMR extends the DMR framework with node embeddings to improve group alignment and relevance, gSTM builds upon STM with additional sparsity controls and group-specific deviations to improve topic coherence and interpretability. These models provide complementary solutions for creating personalized and scalable support groups in online healthcare forums.

\subsection{Incorporating Node Embeddings}

A key innovation in our proposed models, gDMR and gSTM, is the integration of node embeddings derived from user interaction data. These embeddings capture latent relationships between users by modeling their interactions within the online forum, such as replies, posts, and mentions. This allows the models to move beyond basic demographic and textual similarities, incorporating deeper, graph-based relational data that enhances support group formation.

\paragraph{Graph Construction} 
We begin by constructing a directed graph \( G \) to represent user interactions, where each node corresponds to a user. The edges between nodes represent interactions, such as comments or replies, extracted from the dataset. The weight of each edge reflects the frequency of interactions between users, with the edge weight increasing by 1 for each additional interaction. We also store the \texttt{post\_id} as an attribute of the edge, allowing us to track the context of each interaction.

\paragraph{Node Embedding via Node2Vec} 

To generate node embeddings, we employ the \textbf{Node2Vec} algorithm \cite{grover2016node2vec}, which has also been used in previous computational group formation work \cite{fang2022matching, fang2022graph}. Node2Vec learns low-dimensional representations of nodes in a graph by simulating random walks. It explores both the local and global structures of the graph by sampling random walks of length 30 for each node and repeating this process 200 times. Each node is embedded into a 64-dimensional space, capturing its relative position and interaction patterns in the network.

\begin{itemize}
    \item \textbf{Graph Input:} The directed graph \( G \) serves as input to Node2Vec, where each node represents a user and each edge represents an interaction between users.
    \item \textbf{Key Parameters:}
        \begin{itemize}
            \item \textbf{Dimensions:} The node embeddings are 64-dimensional vectors, capturing detailed interaction-based information about users in the graph.
            \item \textbf{Random Walks:} For each user, 200 random walks of length 30 are performed, enabling the algorithm to capture both local and broader structural information.
            \item \textbf{Window Size:} The context window size is set to 10, which controls the range of neighboring nodes considered when learning the embedding of a given node.
        \end{itemize}
\end{itemize}

\paragraph{Embedding Usage in gDMR} 
In the \textbf{gDMR model}, the learned node embeddings are integrated as additional covariate features into the regression component. Specifically, the embeddings complement the demographic and textual data by providing graph-based relational insights. These enhanced features allow gDMR to better model group membership weights \( \alpha_{ug} \), improving its ability to form contextually coherent and demographically relevant support groups.

\paragraph{Embedding Usage in gSTM} 
In the \textbf{gSTM model}, node embeddings are incorporated as structured covariates into the group membership mean \( \mu_{u,g} \). The embeddings interact with the regression coefficients \( \lambda_g \) to inform group assignments, ensuring that both interaction-based relationships and demographic features influence user membership proportions \( \theta_u \). By leveraging these embeddings, gSTM improves its capacity to model fine-grained group-level topic deviations, resulting in enhanced topic coherence and interpretability.

The integration of node embeddings in both gDMR and gSTM models enables the incorporation of structural information derived from user interactions. While gDMR uses the embeddings to refine group membership weights through exponential regression, gSTM leverages them to inform group membership means and deviations, enhancing flexibility in topic modeling. Together, these models provide a robust solution for automating the formation of personalized and contextually coherent support groups.

\begin{figure}[t]
        \centering
        \includegraphics[width=\linewidth]{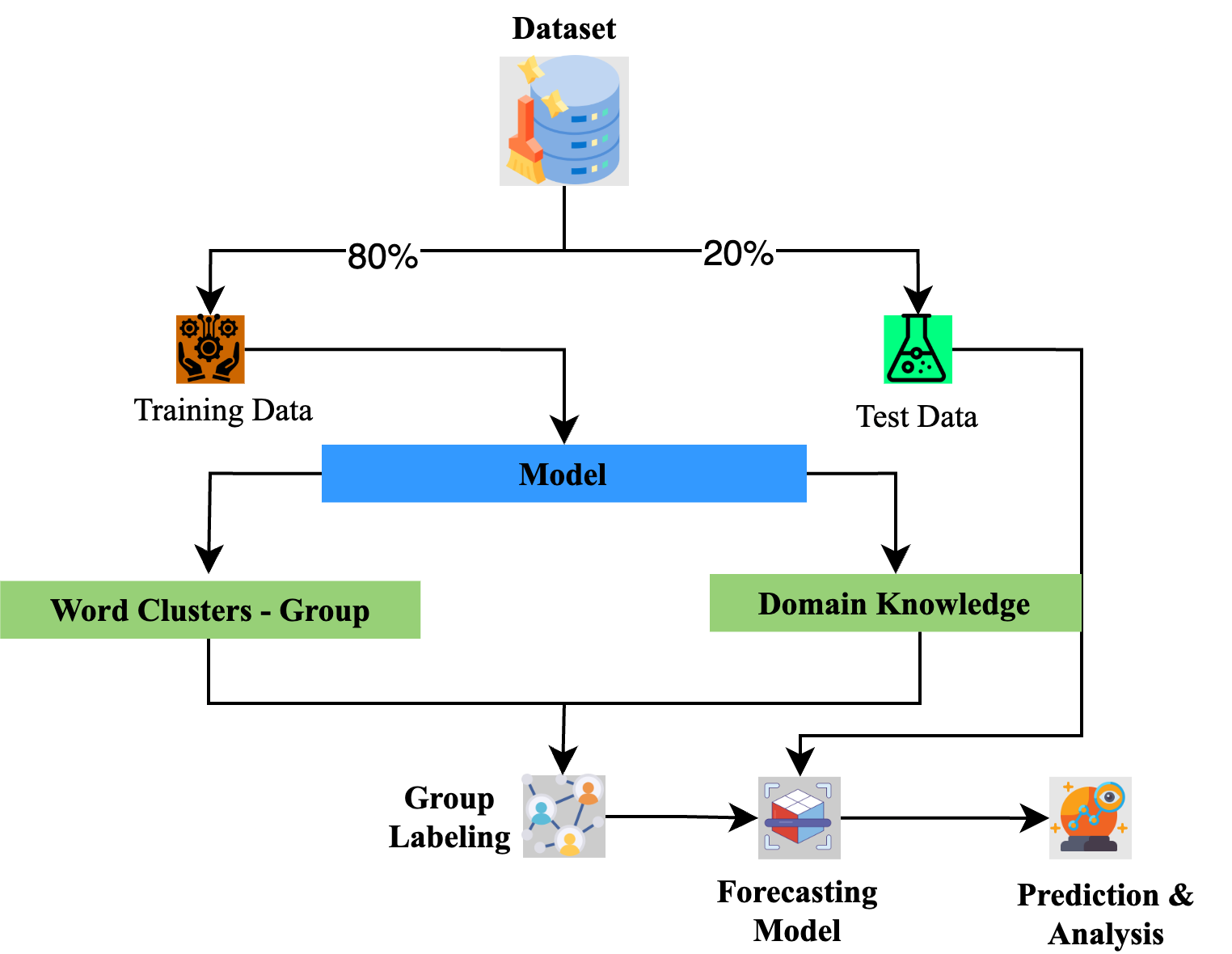}
        \caption{Model Methodology Flowchart.}
        \label{fig:model_flowchart}
\end{figure}

\begin{figure}[t]
        \centering
        \includegraphics[width=\linewidth]{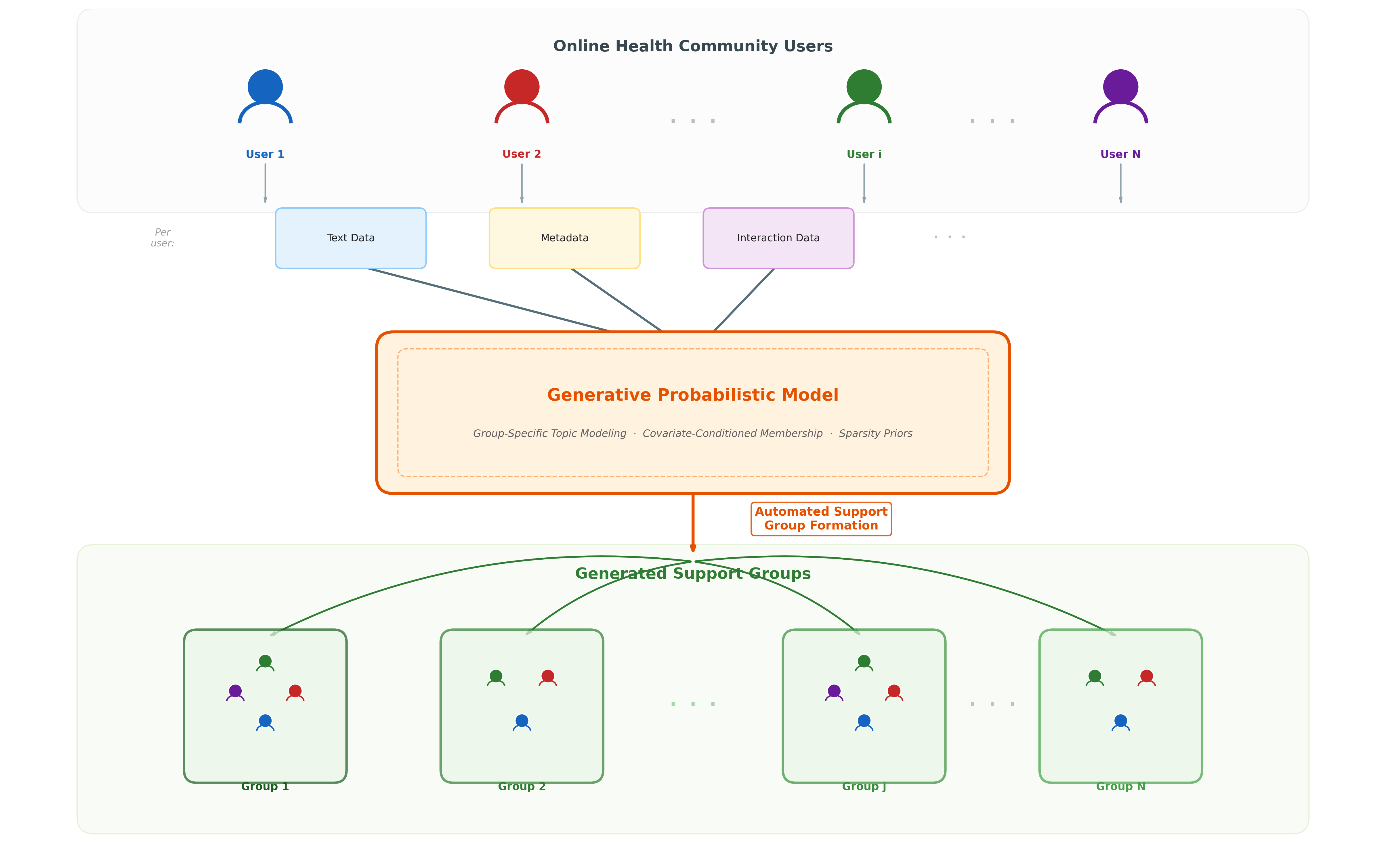}
        \caption{Support Group Formation Flowchart.}
        \label{fig:support_flowchart}
\end{figure}

\subsection{Training Procedure}

In this research, we experiment separately with two distinct topic modeling frameworks—the gDMR and the gSTM—to automate the formation of personalized support groups. Each model independently leverages textual content, demographic characteristics, and user interaction data to identify contextually relevant and coherent user groups within online health forums. The overall methodological approach is depicted in Figures~\ref{fig:model_flowchart} and~\ref{fig:support_flowchart}.

To ensure robustness and stability in group discovery and topic coherence, distinct training strategies tailored to the specific assumptions and computational requirements of each model are employed. Although both models utilize the same underlying features, their parameter initialization, inference methods, and optimization procedures differ significantly. The performance and outcomes of the two models are subsequently compared through comprehensive evaluations, highlighting their relative strengths and applicability in forming personalized support groups.

\subsubsection{gDMR Training Procedure}

The training procedure for the gDMR model integrates topic modeling with user-specific metadata to form personalized and interpretable support groups. This process adopts a multi-phase strategy, leveraging a warm-start initialization followed by iterative optimization, with hyperparameters carefully tuned to ensure robust performance. Specifically, the model was configured with the following hyperparameters: number of groups \( G = 20 \), Dirichlet prior for group-word distributions \( \beta = 0.01 \), and regression variance parameter \( \sigma = 1.0 \). The training spanned a total of 1000 Gibbs iterations, divided into distinct phases to balance semantic coherence and user-specific refinement.

\paragraph{Warm-Start Initialization}
The training commenced with a warm-start phase, employing standard Latent Dirichlet Allocation (LDA) \cite{blei2003latent} for the first 700 iterations. This initialization relied solely on document-level word co-occurrence patterns to establish coherent initial topic assignments. By providing a stable foundation, this phase facilitated the subsequent integration of demographic and interaction-based features, enhancing convergence speed and model stability. Further details on the warm-start implementation are provided in Appendix \ref{sec:appendixA}.

\paragraph{Iterative Optimization Phase}
Following the warm-start, the training transitioned to an iterative optimization phase from iteration 701 to 1000. During these 300 iterations, the model alternated every 10 Gibbs iterations between two key processes: collapsed Gibbs sampling for updating group assignments and Broyden–Fletcher–Goldfarb–Shanno (BFGS) optimization for refining regression parameters \cite{griffiths2004finding}. 

\begin{itemize}
    \item \textbf{Collapsed Gibbs Sampling}: Group assignments \( Z^{(i)} \) were updated according to the posterior probability:
    \[
    P(Z^{(i)} = g | Z^{(-i)}, W) \propto \frac{(n_{g,w}^{(-i)} + \beta_w)(n_{u,g}^{(-i)} + \alpha_{ug})}{(n_g^{(-i)} + W\beta)}
    \]
    where \( n_{g,w}^{(-i)} \) denotes the count of word \( w \) in group \( g \) (excluding the current word), \( n_{u,g}^{(-i)} \) represents the number of words from user \( u \) assigned to group \( g \) (excluding the current word), \( \beta_w \) is the Dirichlet prior for word distributions, and \( \alpha_{ug} \) is the regression parameter linking users to groups.

    \item \textbf{BFGS Optimization}: Concurrently, regression parameters \( \lambda \) and \( \gamma \) were optimized using BFGS to maximize the log-posterior objective function:
    \begin{equation*}
    \begin{aligned}
    l(\lambda, \gamma) &\propto \sum_u \left( \log \Gamma\left( \sum_i \exp(\mathbf{x}_u^T \lambda_i) + \gamma_{g_i} \right) \right. \\
    &\quad \left. - \log \Gamma\left( \sum_i \exp(\mathbf{x}_u^T \lambda_i) + \gamma_{g_i} + n_u \right) \right) \\
    &\quad + \sum_u \sum_i \left( \log \Gamma \left( \exp(\mathbf{x}_u^T \lambda_i) + \gamma_{g_i} + n_{i|u} \right) \right. \\
    &\quad \left. - \log \Gamma\left( \exp(\mathbf{x}_u^T \lambda_i) + \gamma_{g_i} \right) \right) \\
    &\quad - \sum_{i,d} \frac{\lambda_{id}^2}{2\sigma^2} - \sum_i \gamma_{g_i}
    \end{aligned}
    \end{equation*}
    Here, \( \mathbf{x}_u \) encapsulates user-specific features (e.g., demographics or node embeddings), \( \lambda_i \) are regression coefficients for the \( i \)-th feature, \( \gamma_{g_i} \) are group-specific adjustments, and \( \sigma^2 \) regularizes the regression coefficients. The full derivation and implementation details are elaborated in Appendix \ref{sec:appendixB}.
\end{itemize}

This dual approach ensured that the gDMR model effectively captured user-group associations while preserving topic coherence, iteratively refining the model to reflect both semantic and metadata-driven structures.

\paragraph{Parameter Stabilization and Convergence}
To achieve parameter stability, a burn-in period was incorporated during the latter stages of training. Although the exact duration varied slightly in prior descriptions, the final iterations (approximately the last 300) allowed the group assignments and regression parameters to stabilize. Convergence was rigorously assessed using log-likelihood metrics and the held-out likelihood, estimated via Annealed Importance Sampling (AIS) \cite{neal2001annealed}. This evaluation confirmed the model’s predictive performance on unseen data, with comprehensive details provided in Appendix \ref{sec:appendixB}.

\subsubsection{gSTM Training Procedure}

The gSTM model leverages a variational Expectation-Maximization (EM) algorithm~\cite{blei2007correlated, ahmed2007seeking, wang2013variational} to infer latent topics while incorporating group-specific structures and metadata. This training procedure iteratively estimates the posterior distributions of latent variables and optimizes model parameters to maximize the Evidence Lower Bound (ELBO). Below, we outline the gSTM training process in a comprehensive and academically rigorous manner.

\paragraph{Variational EM Algorithm}

The gSTM employs a fast variant of the variational EM algorithm to fit the model to textual data. This algorithm alternates between two steps to maximize the ELBO, defined as:

\[
\mathcal{L} = \mathbb{E}_q[\log P(W, Z, \eta, \kappa | \gamma, \Sigma)] - \mathbb{E}_q[\log q(Z, \eta, \kappa)]
\]

The variables and parameters are:
\begin{itemize}
    \item \( W \): Observed word data from documents.
    \item \( Z \): Latent topic assignments for individual words.
    \item \( \eta \): Latent variables representing document-level topic proportions, modeled via a logistic normal distribution.
    \item \( \kappa \): Group-specific deviations in topic content, influenced by covariates, treated as latent variables with priors.
    \item \( \gamma \): Parameters mapping document metadata to topic prevalences.
    \item \( \Sigma \): Covariance matrix capturing topic correlations.
\end{itemize}

The ELBO balances the expected log-likelihood of the data and the Kullback-Leibler (KL) divergence between the variational distribution \( q(Z, \eta, \kappa) \) and the true posterior. Here, \( \gamma \) and \( \Sigma \) are parameters optimized during training, while \( Z \), \( \eta \), and \( \kappa \) are latent variables with approximated posteriors.

\begin{enumerate}
    \item \textbf{E-Step}: The variational distribution \( q(Z, \eta, \kappa) \) is optimized to approximate the posterior \( P(Z, \eta, \kappa | W, \gamma, \Sigma) \) by minimizing the KL divergence, updating variational parameters based on current \( \gamma \) and \( \Sigma \) estimates.
    \item \textbf{M-Step}: The parameters \( \gamma \) and \( \Sigma \) are updated to maximize the ELBO, refining metadata’s influence on topic prevalences (\( \gamma \)) and topic correlations (\( \Sigma \)).
\end{enumerate}

This process iterates until convergence, determined by a maximum iteration limit or ELBO stabilization.

\paragraph{Enhancing Interpretability}

To improve interpretability, the gSTM model incorporates sparsity-inducing priors, ensuring that group-level deviations are meaningful and semantically coherent. Additionally, the model summarizes topics using FREX scoring \cite{roberts2016model}, which balances word frequency and exclusivity to provide intuitive representations. FREX labels topics based on the harmonic mean of word probability under the topic and exclusivity to the topic, producing semantically insightful summaries \cite{bischof2012summarizing}.

This iterative process ensures that gSTM captures fine-grained variations in group-level topics while maintaining high topic coherence and interpretability, making it particularly well-suited for analyzing personalized support groups in online health forums.

\paragraph{Practical Implementation}

The gSTM training was implemented with:
\begin{itemize}
    \item \textbf{Number of Groups (\( G \))}: 20, setting topic granularity.
    \item \textbf{Initialization}: Spectral initialization for robust starting values.
    \item \textbf{Maximum EM Iterations}: 75, balancing efficiency and convergence.
\end{itemize}

\subsection{Qualitative Analysis and Sampling}

For each support group generated by our models (gDMR and gSTM), we performed stratified random sampling of 20 users (total group user counts range from approximately 20 to 80). This sample size was chosen to balance representativeness with manual coding feasibility.

\paragraph{Data Extraction}  
All posts authored by the selected users were compiled into datasets serving as the basis for qualitative thematic analysis.

\paragraph{Coding Procedure}  
Qualitative coding was conducted through a rigorous thematic analysis approach \cite{braun2006using, clarke2017thematic}, involving:

\begin{enumerate}
    \item \textbf{Codebook Development:} We constructed an initial codebook based on established health-related themes (deductive codes) drawn from prior literature \cite{chapman2015qualitative}, augmented by emergent themes (inductive codes) identified during preliminary data exploration.
    
    \item \textbf{Independent Coding:} A primary coder systematically applied the coding framework to the posts, refining codes iteratively as new themes emerged. 
    
    \item \textbf{Interrater Reliability:} To ensure rigor, a second coder independently coded a randomly selected 10\% subset of the data. Interrater agreement was quantified using Cohen’s Kappa, yielding a substantial agreement of 0.78.
    
    \item \textbf{Thematic Aggregation:} Codes were aggregated to identify dominant themes, which were then compared with model-derived keywords and representative posts to assess semantic coherence and relevance.
\end{enumerate}


\section{Evaluation Methods}

\subsection{Quantitative Evaluation and Refinement}
We comprehensively evaluated both the gDMR and gSTM models using quantitative and qualitative methods to ensure robust assessment of topic quality and support group coherence.

\paragraph{Log-Likelihood Estimation.}  
To quantitatively assess predictive performance and model fit, we employed Annealed Importance Sampling (AIS)~\cite{neal2001annealed} to estimate the log-likelihood of held-out data. AIS provides reliable estimates by transitioning through intermediate distributions to approximate the posterior accurately, reflecting the models' generalization capabilities.

\paragraph{Topic Coherence.}  
Semantic interpretability and coherence of generated topics were assessed using the UMass coherence metric~\cite{mimno2011optimizing, rosner2014evaluating}. This metric calculates coherence based on the co-occurrence statistics of top representative words within each topic from a reference corpus, where higher scores indicate better topic quality.

\paragraph{Within-Group Similarity.}  
Semantic coherence within groups was quantified by calculating the cosine similarity among textual embeddings (e.g., BERT embeddings)~\cite{reimers2019sentence} of group members' posts. Higher within-group similarity scores reflect greater internal semantic consistency, validating the interpretability of automatically generated support groups.

\subsection{Qualitative Coding Procedure}

To complement the quantitative evaluation, we conducted a qualitative analysis to assess the interpretability of topics and automatic support group formation. For each support group generated by our models (gDMR and gSTM), we randomly selected 20 users. All posts from these selected users were then extracted for detailed manual review and coding.

\subsubsection{Selection and Data Extraction}

For each support group, a stratified random sampling approach was employed to select 20 users, ensuring that the sample was representative of the group's overall composition. All posts authored by these selected users were compiled into datasets that served as the foundation for qualitative analysis.

\subsubsection{Coding Methodology}

Qualitative coding was performed systematically by a single expert coder through the following structured approach:

\begin{enumerate}
    \item \textbf{Codebook Development:} An initial coding framework was established based on a thorough review of relevant literature and a preliminary examination of the data. This framework integrated deductive codes derived from established health themes, as well as inductive codes that emerged organically during initial data exploration~\cite{clarke2017thematic}.

    \item \textbf{Independent Coding:} The coder independently reviewed each user's posts in full, meticulously applying the initial coding scheme. During this phase, emergent themes identified, documented, and integrated into the evolving codebook.

    \item \textbf{Thematic Aggregation:} After coding all posts, codes were aggregated to identify dominant themes within each support group. These emergent themes were systematically compared with the top keywords and representative posts generated by the models to assess semantic coherence and practical relevance.
\end{enumerate}


\section{Experimental Results}


\subsection{Support Group Formation}

In addition to the probabilistic topic models (gDMR and gSTM), we implemented a multi-stage post-processing approach to form coherent and practical support groups by combining textual content and user attribute data. This procedure is designed to refine the raw model outputs into well-balanced, actionable peer groups suitable for online health forums.

\subsubsection{Initial Group Assignment via gDMR}

The initial step leverages the gDMR model to generate high-level group assignments for users. This model integrates demographic information, node embeddings from user interaction networks, and textual content similarity to produce personalized cluster memberships.

\paragraph{Similarity Computation}

Textual similarity was quantified using TF-IDF vectorization of user posts, capturing semantic relationships around key health themes (e.g., chronic illness, mental health). Simultaneously, feature similarity was calculated from user embeddings (derived from interaction data) and one-hot encoded demographic features such as gender, age group, country, and membership status.

A weighted combination scheme was applied: 70\% weight was assigned to textual similarity and 30\% to feature similarity, generating a composite similarity score for each user-group pairing. Users were assigned to the group for which this combined similarity was maximal, aligning both content and demographic affinities.

\subsubsection{Refinement via Constrained K-Means Clustering}

To improve internal group coherence and achieve manageable group sizes for real-world application, we applied a constrained K-Means clustering algorithm to users within each initial group.

\paragraph{Constraints and Rationale}

Clusters were constrained to sizes between 10 and 30 members, balancing the trade-off between group cohesion and feasibility for meaningful peer interaction. These size constraints are informed by literature on optimal support group sizes for engagement and manageability \cite{chen2012exploring}. Features used in clustering were standardized with \texttt{StandardScaler} to ensure consistent scaling across heterogeneous attributes.

\subsubsection{Evaluation Framework}

\paragraph{Comparison Baselines}

To rigorously evaluate the semantic coherence of the formed groups, we compared the support groups generated by our models against a baseline of randomly assigned groupings. Although the dataset contains organic MedHelp forums and user-created groups, these were not used as baselines due to data availability and structural differences. Incorporating such organic groupings in future work would provide valuable comparative insight.

\paragraph{Semantic Similarity Analysis}

\begin{figure}[t]
    \centering
    \includegraphics[width=\linewidth]{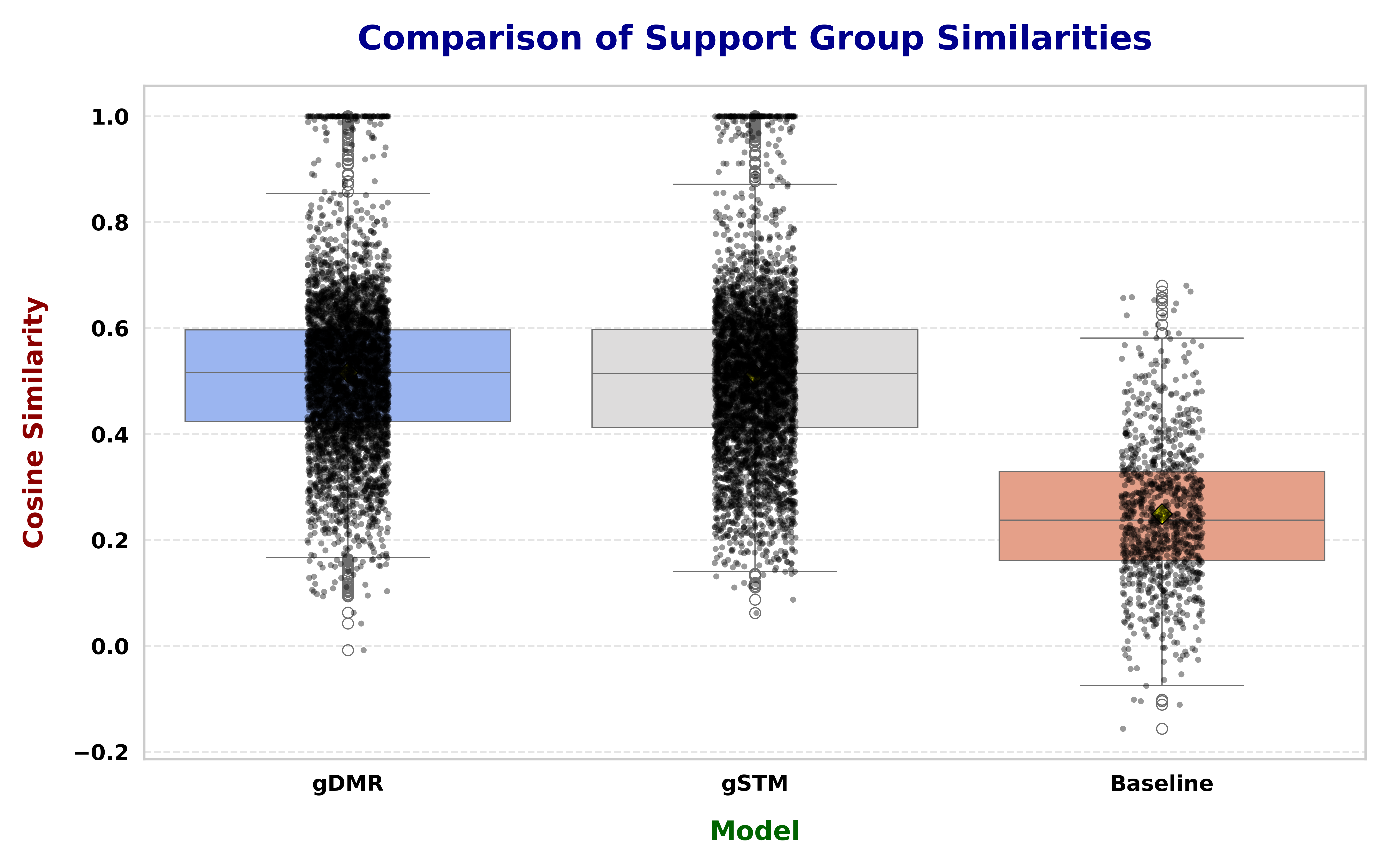}
    \caption{Comparison of Within-Group and Random Baseline Similarities, highlighting enhanced semantic coherence within groups formed by gDMR and gSTM models.}
    \label{fig:similarity_comparison}
\end{figure}

Within-group semantic coherence was measured using cosine similarity of TF-IDF vectors representing user posts. Figure~\ref{fig:similarity_comparison} illustrates the distribution of these similarities across groups formed by gDMR, gSTM, and the random baseline.

\subsubsection{Results and Interpretation}

\begin{itemize}
    \item \textbf{gDMR:} Achieved a median cosine similarity of approximately 0.55 with an interquartile range (IQR) of 0.42 to 0.6, indicating generally strong thematic alignment, though with some variability.
    \item \textbf{gSTM:} Exhibited comparable median similarity (~0.55) but with a tighter IQR and fewer low-similarity outliers. Notably, gSTM produced groups with exceptionally high similarity ($>0.85$), reflecting highly coherent clusters.
    \item \textbf{Random Baseline:} Displayed significantly lower median similarity (~0.27) and broader variability, confirming the robustness of the model-based groupings.
\end{itemize}

These results highlight that both gDMR and gSTM generate support groups with substantially greater semantic coherence than random assignment, validating their practical utility.

\subsubsection{Key Insights}
Despite comparable median performances, gSTM exhibits greater internal consistency and reduced variability, indicating stronger reliability and coherence.

Given its superior consistency and interpretability, gSTM is particularly suited to real-world applications demanding precise thematic grouping and meaningful peer interactions.

We evaluated the proposed gDMR and gSTM models by comparing their performance against several baseline models: LDA, DMR, and STM. To comprehensively assess the quality of the generated support groups and topics, we report three primary performance metrics: perplexity, held-out log-likelihood, and topic coherence. These metrics collectively evaluate the models’ predictive capability on unseen data and the semantic interpretability of the derived topics.

\subsection{Experimental Setup}

For all experiments, we partitioned the dataset into training and testing sets using an 80/20 split. This approach allows us to evaluate the models' ability to generalize beyond the training data, which is critical for applications such as support group formation where new user data continuously emerges.

The 80/20 split balances sufficient training data for model learning with an adequate hold-out set for reliable performance estimation.

\subsection{Quantitative Performance Evaluation}

Table~\ref{tab:log_likelihood} summarizes the comparative performance of all models in terms of held-out log-likelihood and topic coherence.

\begin{table*}[t]
\centering
\begin{tabular}{lll}
\toprule
\textbf{Model} & \textbf{Held-out Log-Likelihood} & \textbf{Coherence Score}\\
\midrule
LDA Model & -606.874 & -4.488 \\
DMR Model & -465.230 & -4.789 \\
gDMR Model (without node embeddings) & -463.778 & -5.048 \\
gDMR Model (with node embeddings) & -403.350 & -2.401 \\
STM Model & -7.669 & -4.499 \\
gSTM Model & -7.314 & -2.165 \\
\bottomrule
\end{tabular}
\caption{Comparison of held-out log-likelihood and coherence scores across models.}
\label{tab:log_likelihood}
\end{table*}

\subsubsection{Held-out Log-Likelihood}

Held-out log-likelihood evaluates each model's ability to predict unseen data, with higher (less negative) values indicating better predictive performance. We estimated this metric using Annealed Importance Sampling (AIS) \cite{wallach2009evaluation}, a robust technique that approximates the partition function for complex probabilistic models.

The gDMR model incorporating node embeddings achieved a held-out log-likelihood of -403.350, which represents an improvement over its variant without node embeddings (-463.778). Although this numerical difference may appear moderate, in the context of log-likelihood metrics for large datasets, such improvements are meaningful and statistically significant given the scale of data and model complexity.

Similarly, the gSTM model outperformed the STM baseline (-7.669 vs. -7.314), demonstrating the benefits of incorporating group-specific structures and sparsity priors. It is important to note that STM and gSTM log-likelihood values are on a different scale due to their model structures and likelihood formulations.

Baseline models such as LDA and DMR showed lower log-likelihoods, reflecting their limitations in capturing complex user interactions and demographic heterogeneity that the proposed models address.


\subsubsection{Topic Coherence}

Topic coherence measures the semantic interpretability of generated topics by assessing the degree of semantic relatedness among top words within each topic. This is critical for ensuring that support groups correspond to meaningful and engaging themes.

As shown in Table~\ref{tab:log_likelihood}, gDMR with node embeddings achieved a coherence score of -2.401, outperforming both the no-embedding variant (-5.048) and the DMR baseline (-4.789). This underscores the value of incorporating graph-based user interaction features to enhance semantic consistency.

The gSTM model attained the best coherence score (-2.165), exceeding that of STM (-4.499). This improvement reflects gSTM’s ability to model fine-grained group-level topic deviations and enforce sparsity, which promotes topic distinctiveness and interpretability.

\subsubsection{Summary of Quantitative Results}

Both gDMR and gSTM demonstrate clear advantages over traditional models in terms of predictive accuracy and topic coherence.  
\begin{itemize}
    \item \textbf{gDMR:} Node embeddings enrich the model with relational context, enhancing predictive and interpretative quality.  
    \item \textbf{gSTM:} Group-specific sparsity and structured covariates improve thematic clarity and topic quality.  
\end{itemize}  
These complementary strengths highlight the potential of integrating textual, demographic, and interaction data to support scalable, personalized group formation in online health communities.

\subsection{Qualitative Performance Evaluation}

We conducted a qualitative analysis comparing model-generated support groups to thematic codes derived from independent qualitative analysis.

\subsubsection{Overview}

For each support group produced by gDMR and gSTM, Table~\ref{tab:gdmr} and Table~\ref{tab:gstm} (Appendix~\ref{sec:appendixC}) present the model’s top keywords alongside representative posts and their corresponding qualitative codes. This comparison provides insight into the alignment between computational clusters and nuanced, context-rich themes identified through qualitative analysis.

\subsubsection{gDMR-Formed Support Groups}

The gDMR model identified 20 distinct groups, each characterized by a dominant theme corresponding closely with qualitative coding. For example:

\begin{itemize}
    \item \textbf{Group 0} focuses on gastrointestinal issues, with keywords such as ``stomach,'' ``eat,'' and ``bowel.'' The majority of posts were coded as relating to digestive disorders, encompassing conditions like diverticulitis and related symptom management.
    \item \textbf{Group 1} centers on cardiac health concerns, with terms like ``heart,'' ``rate,'' and ``chest.'' Posts often addressed symptoms such as palpitations and chest pain, coded under cardiac conditions or health anxiety specific to cardiac symptoms. (Note: health anxiety differs from cardiac health in that it reflects psychological concern about symptoms rather than medically diagnosed conditions.)
    \item \textbf{Group 5} primarily includes dermatological terms (e.g., ``skin,'' ``red,'' ``bump'') but also captures overlapping respiratory concerns and anxiety symptoms. This group illustrates how the model clusters users experiencing somatic symptoms that span multiple health areas, such as respiratory discomfort accompanied by anxiety about physical sensations.
\end{itemize}

These examples demonstrate that the gDMR model clusters posts by the dominant health concern or symptom focus, aligning with how real-world support groups form around shared experiences. For instance, users describing post-nasal drip and shortness of breath (initially coded under respiratory issues) clustered with others experiencing somatic health anxieties, suggesting meaningful grouping by user concern rather than strictly clinical categories.

Moreover, demographic attributes played a significant role:  
- The Women’s Health group (Group 8) predominantly comprised women discussing menstruation and pregnancy-related topics, supported by high gender covariate weights.  
- The HIV risk anxiety group (Group 7) skewed toward male users, reflecting the gendered nature of the content and user base.  
- Age-related patterns emerged in groups addressing parenting and sleep disturbances, mirroring real-world demographic trends.

\subsubsection{gSTM-Formed Support Groups}

The gSTM model generated groups with strong thematic coherence and clear alignment to qualitative codes:

\begin{itemize}
    \item \textbf{Group 0} (Parenting \& Behavioral Issues) featured keywords like ``child,'' ``son,'' and ``baby,'' matching posts coded as behavioral concerns in children.
    \item \textbf{Group 5} (Cardiac Health) clustered posts focused on heart symptoms and diagnostics, reflecting precise thematic grouping.
    \item \textbf{Group 14} (Forum Meta-Discussion) clustered posts related to community management rather than health topics, showcasing the model’s ability to capture meta-level themes.
\end{itemize}

Demographic and geographic covariates further enriched group characterization, revealing age and country-specific participation patterns consistent with health topics.

\subsection{Forum-Level Analysis: From Large Forums to Personalized Support Groups}

In online health communities, a \textit{forum} typically refers to a broad topical community where all users discuss a shared health condition or concern (e.g., ``Stroke Help'' or ``STDs Help''). In contrast, a \textit{support group} denotes a smaller, more focused cluster of users grouped according to highly similar needs, experiences, or contexts, as discovered by our computational models.

To evaluate the practical effectiveness of our approach, we conducted a comparative analysis of five randomly selected MedHelp forums: \textbf{Stroke Help}, \textbf{Back \& Neck Help}, \textbf{STDs Help}, \textbf{Ovarian Cancer Help}, and \textbf{Orthopedics Help}. Each of these forums originally consisted of a large, heterogeneous user base---for instance, the Stroke Help forum contained 52 unique users with a forum-level median semantic similarity of 0.407, while Back \& Neck Help included 1459 users (similarity 0.356).

Both the gDMR and gSTM models automatically partitioned these broad forums into multiple, smaller support groups, each exhibiting substantially higher within-group semantic similarity and increased thematic specificity. For example, in the \textbf{Stroke Help} forum, gDMR identified support groups such as ``Support\_group19-431'' (2 users, median similarity 0.569), ``Support\_group14-140'' (2 users, 0.560), and ``Support\_group2-25'' (3 users, 0.461), grouping users with shared experiences such as caregiver concerns or post-stroke rehabilitation. The gSTM model produced groups such as ``Support\_group11-10'' (4 users, 0.513) and ``Support\_group11-142'' (2 users, 0.566), achieving similar or higher thematic coherence.

This trend was consistent across all forums. In \textbf{Back \& Neck Help} (1459 users, forum similarity 0.356), both models formed support groups with internal median similarities often above 0.60, significantly surpassing the baseline similarity of the full forum. In the \textbf{STDs Help} forum (3376 users, similarity 0.430), gDMR identified groups like ``Support\_group11-4'' (35 users, 0.556) and ``Support\_group5-434'' (30 users, 0.598), while gSTM generated ``Support\_group2-120'' (28 users, 0.606) and ``Support\_group12-68'' (28 users, 0.547), each with high thematic cohesion focused on specific aspects such as recent diagnoses or risk assessment.

Similarly, in \textbf{Ovarian Cancer Help} (564 users, similarity 0.391) and \textbf{Orthopedics Help} (319 users, 0.273), both models produced numerous support groups with within-group median similarities up to 0.590 (Ovarian Cancer Help, gDMR) and 0.479 (Orthopedics Help, gSTM), with group sizes ranging from small dyads to larger, thematically consistent clusters.

In summary, both quantitative (semantic similarity) and qualitative (thematic focus) analyses demonstrate that our model-generated support groups are markedly more cohesive, personalized, and effective than the original, broad forums. These results underscore the value of automated support group formation for enhancing the granularity and relevance of peer support in large-scale online health communities.

\section{Discussion}

This study introduces two complementary models, the Group-specific Dirichlet Multinomial Regression (gDMR) and the Group-specific Structured Topic Model (gSTM), designed to enhance automated support group formation within online health communities (OHCs). Both models significantly advance current methodologies by integrating demographic metadata, user interaction features, and user-generated textual content. Collectively, these models address critical limitations identified in traditional, manually created support groups, which typically lack personalization, scalability, and interpretability \cite{eysenbach2004, breuer2015}.

The gDMR model extends the Dirichlet Multinomial Regression framework by incorporating node embeddings and group-specific parameters, effectively capturing latent relational structures among users. Our experiments demonstrate that this extension significantly improves predictive accuracy (held-out log-likelihood) and semantic coherence compared to baseline models such as DMR, STM, and LDA. Specifically, the incorporation of interaction-based node embeddings enables gDMR to accurately identify and leverage complex relational patterns and demographic nuances, enhancing the formation of personalized and contextually relevant support groups.

Complementarily, the gSTM model provides a rigorous framework for fine-grained thematic analysis through structured covariates, sparsity-inducing priors, and group-specific deviations. Our results indicate that gSTM surpasses baseline models in generating semantically coherent and interpretable topics. The hierarchical nature of gSTM allows it to capture subtle thematic variations across groups, particularly beneficial when clear topic interpretability is essential for practical applications such as targeted health interventions or personalized information dissemination \cite{roberts2013structural}.

\subsection{Comparative Insights}

This work offers valuable advances for multiple stakeholders involved in managing and participating in OHCs, including platform designers, healthcare practitioners, patient advocacy groups, and researchers in digital health informatics. These audiences stand to benefit from improved methods for fostering personalized, dynamic, and meaningful peer support that better reflect the complex realities of users’ health experiences.

Both gDMR and gSTM substantially outperform traditional static categorization approaches, which remain prevalent in popular platforms such as Reddit and Facebook. These conventional methods typically rely on predefined, rigid categories or community labels—for example, subreddits organized strictly around specific diagnoses (e.g., r/diabetes, r/depression) or Facebook groups manually created for discrete health conditions. Such static structures inherently limit adaptability, failing to capture the heterogeneity of individual user journeys, the evolving nature of health concerns, and the nuanced social interactions occurring within these communities \cite{burlingame_2004, greene2011}.

In practice, static categorization manifests as rigid forum sections or fixed group memberships that do not adjust to changing user needs or intersecting health issues. Users with overlapping or multiple conditions may be forced to engage in multiple disjointed groups, hindering holistic support. Moreover, these categories rarely incorporate user interaction patterns or demographic context, limiting their capacity to foster meaningful connections or personalize support.

In contrast, gDMR and gSTM dynamically integrate demographic metadata and user interaction networks with textual content, enabling adaptive, nuanced support group formation that better aligns with individual and collective experiences. gDMR excels at relational clustering through node embeddings, effectively capturing social connectivity and fostering peer groups that mirror natural interaction patterns. Meanwhile, gSTM enhances thematic specificity and interpretability by modeling fine-grained topic variations, enabling the formation of groups with clearly distinguishable and relevant health concerns.

Together, these models provide a comprehensive framework addressing both social and thematic dimensions critical to effective support in OHCs. For platform developers and healthcare facilitators, this translates into the ability to deliver tailored peer support networks that evolve with users’ health trajectories, promote sustained engagement, and ultimately improve patient outcomes through better community cohesion and information relevance.

\subsection{Addressing Bias and Fairness}

Automated support group formation, while offering scalability and personalization, raises important ethical concerns around bias and fairness due to its reliance on demographic attributes and interaction-based features \cite{hashmi2022}. Attributes such as age, gender, and geographic location are critical for tailoring support, yet overreliance on these factors risks perpetuating existing social inequities. For instance, underrepresented groups—such as racial minorities or socioeconomically disadvantaged populations—may be systematically marginalized if the models preferentially cluster users based on dominant demographic patterns, thereby limiting their access to effective peer support.

Moreover, node embeddings employed in gDMR, which encapsulate user interaction patterns, may inadvertently reflect biases present in the underlying social networks. For example, if certain groups engage less frequently or are socially isolated within the online community, their sparse connections can result in lower-quality embeddings, and consequently, less favorable group assignments.

To concretely illustrate, consider a hypothetical online health forum where younger users dominate discussions and form tightly connected clusters, while older adults participate less frequently and have weaker interaction networks. Without fairness-aware adjustments, the model might consistently assign older adults to less coherent or smaller groups, diminishing their access to supportive peer networks.

Addressing such biases requires explicitly integrating fairness constraints into the group formation process. This could involve ensuring demographic parity—where each protected group (e.g., age, gender, ethnicity) receives equitable representation across formed groups—or optimizing for equality of opportunity by guaranteeing comparable support quality metrics for all user segments.

Practically, this necessitates developing fairness-aware extensions of gDMR and gSTM that incorporate such constraints during optimization or as post-processing corrections. Additionally, ongoing evaluation using fairness metrics should accompany traditional performance measures to monitor and mitigate disparate impacts.

Finally, transparent decision-making frameworks that allow users and community managers to understand and contest group assignments can further promote inclusivity and trust.

By embedding fairness considerations into the technical and operational pipeline of automated support group formation, future models can better uphold ethical standards and ensure that digital health communities serve the diverse needs of all users effectively.

\section{Limitations, Ethical Considerations, and Future Directions}

While the proposed gDMR and gSTM frameworks advance automated support group formation, several important limitations and ethical considerations warrant attention in future research to ensure robust, fair, and scalable deployment.

\subsection{Data Quality and Representation}

The models’ effectiveness critically depends on the quality, diversity, and representativeness of input data. Imbalances in demographic groups or limited variability in health conditions can reduce generalizability and unintentionally reinforce systemic biases. Ensuring equitable outcomes—in this context, fair and unbiased representation and access to supportive peer groups across all demographic and social segments—is essential. Future work should emphasize comprehensive, balanced data curation and incorporate fairness-aware modeling strategies to mitigate disparities.

\subsection{Computational Scalability and Real-Time Adaptation}

gSTM’s complexity, due to structured priors and topic correlations, poses computational challenges for training and inference, particularly on large-scale or resource-constrained platforms. Enhancing computational efficiency through algorithmic optimization, approximate inference methods, or distributed computing will be critical for real-time or near-real-time group adaptation. Incorporating user feedback and evolving behavioral patterns dynamically will further improve group relevance and engagement.

\subsection{Evaluation Beyond Quantitative Metrics}

Traditional evaluation metrics such as log-likelihood and topic coherence effectively measure predictive accuracy and semantic quality but fall short of capturing user experience and real-world impact. Incorporating user-centric evaluations—such as satisfaction surveys, longitudinal engagement analysis, and qualitative feedback—is vital for assessing practical utility and informing iterative model improvements.

\subsection{Privacy Protections}

Given the sensitive nature of health data, maintaining rigorous privacy safeguards is paramount. Although this study utilized anonymized and de-identified datasets compliant with existing ethical standards, future deployments must adhere to evolving global privacy frameworks (e.g., GDPR). Techniques such as differential privacy or federated learning could be explored to enhance data security and user trust without compromising model performance.

\subsection{Mitigating Bias and Ensuring Inclusivity}

Automated group formation systems must continuously monitor and address potential biases stemming from demographic attributes and interaction patterns. Equitable access here means that users from all backgrounds—regardless of age, gender, ethnicity, or social connectivity—should have fair opportunities to be assigned to supportive and coherent groups. Achieving this requires embedding fairness constraints into model training and evaluation, transparent decision-making mechanisms to explain group assignments, and ongoing assessment of group composition relative to inclusivity benchmarks.

\subsection{Broader Applicability and Future Research}

Beyond healthcare, the flexible nature of gDMR and gSTM suggests applicability in diverse domains such as education (personalized study groups), customer support (dynamic client clustering), and social networking. Future work should explore these contexts while addressing the challenges above.

In summary, by proactively integrating fairness, privacy, scalability, and user-centered evaluation into the development pipeline, future iterations of these models hold promise for transforming how personalized support communities are created and sustained across digital platforms.


\appendix

\section{Appendix A: Justification for Using LDA as a Baseline}
\label{sec:appendixA}

In this study, LDA was chosen as the baseline model due to its established role as a standard topic modeling algorithm in the literature \cite{blei2003latent, blei2012probabilistic}. LDA's widespread use in uncovering latent topics within large text corpora makes it an appropriate reference point for evaluating the performance of more complex models, such as gDMR.

\subsection{Rationale for Using LDA in Initial Iterations}

We initiated the training process with 700 iterations of LDA to provide a stable foundation for topic discovery before introducing the demographic features and regression parameters in gDMR. The use of LDA during the initial phase serves several purposes:
\begin{itemize}
    \item \textbf{Robust Initialization:} LDA effectively identifies coherent topics based solely on textual data, providing a robust starting point for the more sophisticated gDMR model. This ensures that the initial topics are meaningful and reduces the computational burden when transitioning to gDMR.
    \item \textbf{Efficient Training:} Introducing gDMR from the start would require the model to simultaneously learn both topic distributions and the influence of demographic features, potentially leading to slower convergence. By allowing LDA to first converge on the text data, the subsequent introduction of gDMR results in more efficient training and more stable results.
\end{itemize}

\subsection{LDA as a Benchmark for gDMR}

LDA serves as an essential benchmark for evaluating the enhancements offered by gDMR. By comparing the perplexity and log-likelihood metrics, we can quantitatively assess the improvements brought by incorporating demographic features into the topic modeling process. The observed reductions in perplexity and improvements in log-likelihood underscore the gDMR model's ability to provide a more nuanced understanding of the data, particularly in capturing user-specific topics informed by demographic variables.

\section{Appendix B: Collapsed Gibbs Sampling and Evaluation Methods}
\label{sec:appendixB}

\subsection{Mathematical Framework and Derivations}
We present the mathematical foundations for the gradient derivations used in the gDMR model's optimization process~\cite{carpenter2010integrating}. The posterior function \( P(\mathbf{z}, \lambda, \gamma) \) leads to the following log-posterior function:

{
\small 
\begin{align*}
l(\lambda, \gamma) \propto & \sum_u \left[ \log \Gamma\left(\sum_i \left( \exp(\mathbf{x}_u^T \lambda_i) + \gamma_{g_i} \right) \right) \right. \\
& \left. - \log \Gamma\left(\sum_i \left( \exp(\mathbf{x}_u^T \lambda_i) + \gamma_{g_i} \right) + n_u \right) \right] \\
& + \sum_i \left[ \log \Gamma \left( \exp(\mathbf{x}_u^T \lambda_i) + \gamma_{g_i} + n_{i|u} \right) \right. \\
& \left. - \log \Gamma \left( \exp(\mathbf{x}_u^T \lambda_i) + \gamma_{g_i} \right) \right] \\
& - \sum_{i,d} \left( \frac{\lambda_{id}^2}{2\sigma^2} + \log \frac{1}{\sqrt{2\pi\sigma^2}} \right) \\
& - \sum_i \gamma_{g_i}.
\end{align*}
}

The digamma function \( \Psi(x) \), as the logarithmic derivative of the gamma function, is used in the gradient calculations:
\[
\Psi(x) = \frac{\Gamma'(x)}{\Gamma(x)}.
\]

\subsection{Partial Derivatives for Optimization}
We derive the gradients with respect to the model parameters \( \lambda \) and \( \gamma \) as follows, which guide the optimization process using BFGS.

\subsubsection{Gradient with respect to \texorpdfstring{$\lambda_{iu}$}{lambda}}

{
\small
\begin{align*}
    \frac{\partial l}{\partial \lambda_{iu}} = & \sum_u x_{iu} \exp(x_u^T \lambda_i) \\
    & \Bigg[ \Psi\left(\sum_i \alpha_i\right) - \Psi\left(\sum_i \alpha_i + n_u\right) \\
    & + \Psi\left(\alpha_i + n_{i|u}\right) - \Psi\left(\alpha_i\right) \Bigg] 
    - \frac{\lambda_{iu}}{\sigma^2}.
\end{align*}
}

\subsubsection{Gradient with respect to \texorpdfstring{$\gamma_{g_i}$}{gamma\_g\_i}}

{
\small
\begin{align*}
    \frac{\partial l}{\partial \gamma_{g_i}} = & \sum_u \left( \Psi\left(\sum_i \alpha_i\right) - \Psi\left(\sum_i \alpha_i + n_u\right) \right. \\
    & \left. + \Psi\left(\alpha_i + n_{i|u}\right) - \Psi\left(\alpha_i\right) \right) - G.
\end{align*}
}

These gradients help optimize the parameters to improve the formation of group-specific distributions and enhance topic coherence.

\subsection{Annealed Importance Sampling for Log-Likelihood Estimation}

We employed AIS to estimate the log-likelihood on held-out data~\cite{wallach2009evaluation}, approximating the partition function between a simple prior and the complex posterior. This method provides a reliable means to evaluate the model's performance in probabilistic frameworks such as gDMR.

We defined intermediate distributions \( \{P_s(z)\}_{s=0}^S \) with:
\[
P_s(z) \propto P(z | \alpha) \cdot P(w | z, \Phi)^{\tau_s},
\]
where \( \tau_s \) is a temperature schedule, gradually transitioning from prior \( P_0(z) \) to posterior \( P_S(z) \).

Markov chain Monte Carlo (MCMC) sampling is employed at each temperature \( \tau_s \), and the importance weight for each sample sequence is computed as:
\[
w_{AIS} = \prod_{s=1}^{S} \left( \frac{P(w | z^{(s)}, \Phi)^{\tau_s - \tau_{s-1}}}{T_s(z^{(s-1)} \rightarrow z^{(s)})} \right),
\]
where \( T_s(z^{(s-1)} \rightarrow z^{(s)}) \) represents the transition probability between states in the Markov chain.

The log-likelihood estimate for the held-out data is derived by averaging the importance weights across several runs:
\[
\log P(w | \Phi, \alpha) \approx \log \left( \frac{1}{M} \sum_{m=1}^{M} w_{AIS}^{(m)} \right).
\]

This process allows us to evaluate how well the model generalizes to unseen data, ensuring robustness in performance assessment.

\section{Appendix C: Thematic Insights from gDMR and gSTM Models}
\label{sec:appendixC}

\begin{longtable}{p{1.5cm}p{2.5cm}p{5cm}p{2.5cm}p{2.5cm}}
\caption{gDMR Model Support Groups with Top Words and Example Posts} \label{tab:gdmr} \\
\toprule
\textbf{Support Group} & \textbf{Top Words} & \textbf{Representative Post} & \textbf{Qualitative Code} & \textbf{Covariates} \\
\midrule
\endfirsthead

\multicolumn{5}{c}{{\tablename\ \thetable{} -- Continued from previous page}} \\
\toprule
\textbf{Support Group} & \textbf{Qualitative Code} & \textbf{Top Words} & \textbf{Representative Post}  & \textbf{Covariates} \\
\midrule
\endhead

\midrule
\multicolumn{5}{c}{{Continued on next page}} \\
\midrule
\endfoot

\bottomrule
\endlastfoot

Group 0 & Gastrointestinal Disorders & eat, stomach, problem, drink, food & ``I was diagnosed with diverticulitis about 10 days ago... lower left side pain...'' & Grenada, Kiribati, France \\
\midrule

Group 1 & Heart Disease and Symptoms & heart, stress, exercise, rate, normal & ``There is a history of heart disease in my family... For the last 6 months I have been experiencing episodes...'' & Lithuania, Vietnam \\
\midrule

Group 2 & Joint Pain and Swelling Syndrome & problem, diagnosis, physical, increase, episode & ``I am trying to find out why occasionally my wrists, fingers and one foot will swell up for no obvious reason...'' & Male \\
\midrule

Group 3 & HPV Testing & question, possible, concern, point, concerned & ``Does anyone know whether gyn's automatically do the HPV test when they do a pap?'' & Malaysia \\
\midrule

Group 4 & Post-Treatment Side Effects & see, body, begin, different, treat & ``My father, 77, has had GERD for years with increasing severity. A biopsy... [we are seeking] relief from post-radiation therapy.'' & Macau \\
\midrule

Group 5 & Respiratory Issues & area, skin, red, small, bump & ``Six weeks ago I started having post-nasal drip and then it felt like it was all in my chest... short of breath...'' & Greece, Female \\
\midrule

Group 6  & CT Scan & surgery, cancer, remove, cyst, breast & ``I had a CAT scan of the abdomen and I wanted to know if it would show a tumor in my stomach...'' & Sri Lanka, Australia \\
\midrule

Group 7 & Unprotected Oral Sex & sex, hiv, sore, oral, risk & ``I've read many threads about oral sex, but none discuss if brushing your teeth after can increase the risk of HIV.'' & Netherlands, United Kingdom \\
\midrule

Group 8 & Pregnancy Anxiety & period, stop, bad, wonder, go & ``I started bleeding just twelve days after my last period... me and my hubby are trying to get pregnant...'' & Age 30-39, Female, Germany \\
\midrule

Group 9 & Chest Pain & throat, cough, chest, ear, lung & ``My chest felt like it was exploding... They diagnosed [me with acute bronchitis after I collapsed].'' & Germany, Australia \\
\midrule

Group 10 & Thyroid Levels Confusion & normal, high, blood, level, low & ``These numbers still confuse me. I'm still being adjusted for my hypothyroid... What are normal T4, T3, TSH levels?'' & Age 50-59, Ukraine, Netherlands \\
\midrule

Group 11 & Genital Health Concerns & penis, herpe, burn, touch, genital & ``This summer I went to the gyno for the first time with what I thought was a bad infection... [could it be herpes]?'' & Hong Kong, Australia, Poland \\
\midrule

Group 12 & Muscle Fasciculations & leg, arm, MRI, neck, muscle & ``I am a 40-year-old female... I had surgery last May and now I have muscle twitching (fasciculations). Is it benign or ALS?'' & Mauritius, Lithuania, Brazil \\
\midrule

Group 13 & Recurrent Pain & pain, right, left, severe, low & ``I am 31 years old and have had multiple areas of discomfort/pain for several months with no answers...'' & Singapore, Djibouti \\
\midrule

Group 14 & Health Anxiety & me, sometimes, night, eye, head & ``I have had headaches for months and I'm worried it might be an aneurysm. I NEED answers...!'' & China, Vietnam, Female \\
\midrule

Group 15  & Diagnostic Uncertainty & doctor, cause, diagnose, give, long & ``My main question at this point: can a sinus infection (that I had on and off) cause these strange neurological symptoms?'' & Mexico, Age 60-69, Iraq \\
\midrule

Group 16 & STI Test Results & test, symptom, result, negative, blood & ``My doctor called yesterday with my blood test results. I tested positive for HSV-1 (which I already knew) and...'' & Portugal, Jordan \\
\midrule

Group 17 & Behavioral Aggression & child, son, home, daughter, school & ``My 4-year-old son has started having severe anger fits that turn violent...'' & Hong Kong, Lithuania, Age 50-59 \\
\midrule

Group 18 & Nutritional Supplements & treatment, state, medical, patient, recommend & ``I've been on lipotropic injections with my doctor... they have worked, but I've seen he has B1/B6 injections. What's the difference?'' & Thailand, China, Taiwan \\
\midrule

Group 19 & Forum Usability & post, life, people, hear, answer & ``Thank you for being open with your thoughts on the new look and feel of the forum. We are looking to...'' & Canada, France, Djibouti \\
\bottomrule
\end{longtable}
\begin{longtable}{p{2.5cm}p{2.5cm}p{5.5cm}p{2.5cm}p{2.5cm}}
\caption{gSTM Model Support Groups with Top Words and Example Posts} \label{tab:gstm} \\
\toprule
\textbf{Support Group} & \textbf{Top Words} & \textbf{Representative Post} & \textbf{Qualitative Code} & \textbf{Covariates} \\
\midrule
\endfirsthead

\multicolumn{5}{c}{{\tablename\ \thetable{} -- Continued from previous page}} \\
\toprule
\textbf{Support Group} & \textbf{Qualitative Code} & \textbf{Top Words} & \textbf{Representative Post} & \textbf{Covariates} \\
\midrule
\endhead

\midrule
\multicolumn{5}{c}{{Continued on next page}} \\
\midrule
\endfoot

\bottomrule
\endlastfoot

Group 0 & Behavioral Issues & child, son, baby, daughter, school, home & ``I have two young daughters and lately my 3-year-old has been saying disturbing things. I'm frustrated and worried…'' & Djibouti, Nepal, Turkey \\
\midrule
Group 1 & Testing Procedures/Results & test, negative, result, positive, hiv, symptom & ``I know it's a little early to worry, but my period is late and the pregnancy test is still negative. I'm anxious…'' & Costa Rica, Grenada, Macedonia \\
\midrule
Group 2 & Sexual Health Concern & herpes, sex, hsv, penis, genital, wart, partner & ``I have genital warts and started using Aldara cream. My husband has been patient but I would like to have a normal sex life…'' & Croatia, Lithuania, Poland \\
\midrule
Group 3 & Joint Pain & pain, leg, arm, leave, right, muscle, neck, hand & ``I have numbness and tingling in hands and feet. Also experiencing bouts of bad painful gas and constipation.'' & Trinidad and Tobago, Kuwait, Bulgaria \\
\midrule
Group 4 & Bypass Surgery Complications & surgery, treatment, remove, doctor, cancer & ``My dad had six bypass surgeries, now experiencing burning in his chest. What tests should be done?'' & Indonesia, Chile, Macedonia \\
\midrule
Group 5 & Cardiac Testing & heart, normal, rate, beat, stress, chest & ``I am 38 female. I exercise regularly but experience pain, shortness of breath, and dizziness. What tests should I ask for?'' & Portugal, Hong Kong, Malta \\
\midrule
Group 6 & Cancer Risk Concerns & cyst, right, leave, show, ultrasound, scan & ``I had my ovaries removed due to cysts. Can I still get ovarian cancer with just remnants left?''  & Vietnam, Malta, Greece \\
\midrule
Group 7 & Psychiatric Drug Side Effects & anxiety, bad, life, sleep, drug, med & ``Depakote caused anger issues, affecting my marriage. Should I discontinue the medication?'' & Mexico, Brazil, Macau \\
\midrule
Group 8 & Liver and Enzymes & blood, liver, normal, high, test, count & ``I have Hepatitis C, my blood work was slightly elevated, but I feel fine. Should I go for treatment or wait?'' & Chile, Oman, Mauritius \\
\midrule
Group 9 & Chest and Respiratory Issues & pain, chest, stomach, cough, throat, doctor & ``I have been experiencing excessive burping, flatulence, and chest pain. Doctors suspect GERD or anxiety…'' & Iraq, Mauritius, Dominican Republic \\
\midrule
Group 10 & Maternal Health & period, night, happen, pregnant, morning, bad & ``I'm 31 weeks pregnant and having trouble sleeping. My stomach cramps when I lie on my side. Should I be concerned?'' & Kiribati, Nigeria, Netherlands, Age 0-9 \\
\midrule
Group 11 & Vision Problems & eye, head, headache, symptom, brain, ear & ``I have worsening myopia and some floaters. Is my high myopia likely to progress into a serious condition?'' & Yugoslavia, Malaysia, Belgium \\
\midrule
Group 12 & Sexual Health & sex, hiv, protect, condom, risk, sore & ``I had protected sex, but the condom slipped slightly. I later saw a panty liner with a stain. How risky was this?'' & Tanzania, Latvia, Ukraine, Thailand, Age 10-19 \\
\midrule
Group 13 & Diet and Nutrition & eat, drink, food, diet, water, vitamin & ``I experience extreme gas, stomach cramps, and irregular bowel movements. Could I have a gluten allergy or other issue?'' & Oman, Kiribati, China \\
\midrule
Group 14 & Community and Support & post, people, question, hear, forum, answer & ``We are working on splitting the forum into two—one for medical weight loss and one for natural methods…'' & Age 0-9, Croatia, Namibia \\
\midrule
Group 15 & Weight Management & weight, lose, loss, lbs, doctor, pound & ``I am pregnant, but my husband insists I should continue dieting. Is it safe to lose weight during the first trimester?'' & Afghanistan, Spain, Poland, Age 0-9 \\
\midrule
Group 16 & Skin Conditions and Symptoms & skin, red, area, bump, small, rash & ``I have red bumps on my arm that don't respond to treatment. Doctors ruled out shingles. What else could it be?'' & Namibia, Afghanistan, UAE \\
\midrule
Group 17 & Thyroid Function and Testing & thyroid, level, normal, tsh, test, symptom & ``I have hypothyroidism and fluctuating TSH levels. What symptoms should I watch out for? Should my dosage be adjusted?'' & Uganda, Germany, Egypt \\
\midrule
Group 18 & Symptoms and Conditions & infection, symptom, pain, doctor, antibiotic & ``I had kidney stones removed but still feel soreness. How long does recovery usually take?'' & Ukraine, Taiwan, Botswana \\
\midrule
Group 19 & Medical Treatment and Professional Care & pain, test, cfid, right, symptom, doctor & ``My wife suffers from cervical pain extending to her hands. She had MRI results indicating compression at C5-C6…'' & India, Lithuania, Iraq, Age 80-89 \\
\bottomrule
\end{longtable}
\section*{Appendix D: Qualitative Coding Codebook}

\begin{longtable}{p{3.3cm} p{7cm} p{5cm}}
\caption{Qualitative Codebook for Thematic Analysis of Online Health Community Posts} \\
\toprule
\textbf{Code Name} & \textbf{Definition/Description} & \textbf{Example Quotation} \\
\midrule
\endfirsthead
\multicolumn{3}{l}{\emph{(Continued from previous page)}} \\
\toprule
\textbf{Code Name} & \textbf{Definition/Description} & \textbf{Example Quotation} \\
\midrule
\endhead
\midrule
\multicolumn{3}{r}{\emph{(Continued on next page)}} \\
\midrule
\endfoot
\bottomrule
\endlastfoot

Chronic Disease Management & Discussion of ongoing, long-term conditions (e.g., diabetes, heart disease, asthma, hypothyroidism) and their management. & ``I was diagnosed with diabetes last year and struggle to keep my sugar under control.'' \\
\midrule

Acute Symptom or Flare & Description of sudden or short-term physical symptoms, illness flare-ups, or episodic health issues. & ``I started having chest pain and shortness of breath last night.'' \\
\midrule

Diagnostic Uncertainty & Expressions of confusion or seeking clarity about symptoms, medical diagnoses, or test results. & ``My test results still don't explain why I'm having these headaches.'' \\
\midrule

Treatment and Medication & Discussion of treatments, medication regimens, procedures, or medical advice (including adherence, changes, or concerns). & ``My doctor prescribed a new antidepressant, but I'm not sure about the side effects.'' \\
\midrule

Side Effects / Complications & Reference to negative effects or complications following treatment, medication, or medical intervention. & ``Since starting radiation, I've had trouble sleeping and more fatigue.'' \\
\midrule

Health Anxiety / Psychological Distress & Expressions of anxiety, worry, or emotional distress related to health, symptoms, or uncertainty. & ``I'm worried this headache is something serious like a tumor.'' \\
\midrule

Peer Support Request & Request for information, advice, shared experiences, or emotional support from other forum members. & ``Has anyone else experienced this? I could use some advice.'' \\
\midrule

Parenting / Family Care & Concerns related to caring for children, family members, or parenting in the context of health. & ``My 4-year-old son has been having trouble sleeping at night.'' \\
\midrule

Gender-Specific / Reproductive Health & Topics specific to gendered health or reproductive concerns (e.g., pregnancy, menstruation, menopause, male sexual health). & ``I'm anxious about my missed period and whether I might be pregnant.'' \\
\midrule

Diet / Nutrition / Supplements & Discussion of dietary habits, nutrition, food allergies, or use of supplements. & ``I started taking vitamin D and changed to a gluten-free diet.'' \\
\midrule

Exercise / Physical Activity & References to exercise routines, physical activity, or rehabilitation. & ``I began walking every morning after my surgery to help with recovery.'' \\
\midrule

Preventive Health / Screening & Mention of health screenings, preventive measures, or proactive health behaviors. & ``I had my annual mammogram last week as part of a regular checkup.'' \\
\midrule

Community / Forum Meta & Discussion about the forum itself, community rules, structure, or requests for technical help. & ``Does anyone know when the forum layout will be updated?'' \\
\midrule

Stigma / Privacy Concerns & Expressed concerns about privacy, stigma, or fear of being judged for sharing personal information online. & ``I'm hesitant to talk about my diagnosis here because I worry about privacy.'' \\
\midrule

Age-Specific Concerns & Issues specific to certain age groups (e.g., elderly, adolescents, pediatric topics). & ``My elderly mother has been losing her memory lately and I'm not sure what to do.'' \\
\midrule

Geographic / Cultural Context & Mentions of location, cultural differences, or healthcare access as relevant to the user’s situation. & ``Is this treatment available in Canada? My doctor wasn't sure.'' \\
\midrule

Emergent / Other & New or unexpected themes not covered by the above codes (to be defined during coding). & ``I found humor really helps me cope with this illness.'' \\
\end{longtable}

\bibliographystyle{ACM-Reference-Format}  
\bibliography{references}  

\end{document}